# Response to comment on "Resolving spatial and energetic distributions of trap states in metal halide perovskite solar cells"

Zhenyi Ni, Shuang Xu and Jinsong Huang*

Department of Applied Physical Sciences, University of North Carolina, Chapel Hill, NC 27599, USA.
*Corresponding author. Email: jhuang@unc.edu

**Abstract:  Ravishankar et al. claimed the drive-level capacitance profiling (DLCP) method cannot resolve trap density along depth direction in perovskites with given thickness, and explained the measured charges to be a consequence of geometrical capacitance and diffusion capacitance. We point out that the trap densities in DLCP method are derived from the differential capacitance at different frequencies, and thus the background charges caused by diffusion and geometry capacitance has been subtracted. Even for the non-differential doping analysis by DLCP, the contribution from diffusion capacitance is shown to be negligible and contribution from geometry capacitance is excluded. Additional experiment results further support the measured trap density represents the actual trap distribution in perovskite solar cells.**

In this commentary, Ravishankar et al. derived a carrier density distribution in semiconductor devices based on capacitance model by assuming the combination of geometrical capacitance ($C_g$) and diffusion capacitance ($C_d$) as the dominant capacitance for semiconductor devices:
$$C = C_g + C_0 \exp(qV/mkT) \qquad (1)$$
This is the equation for all analysis in the commentary. The authors derived the carrier density by assuming this apparent (total) capacitance under forward bias follows the equations developed for reverse bias capacitance in diode (equation S2 in commentary). It is unknown whether this assumption stands, because neither geometrical nor diffusion capacitance is a function of carrier concentration. It is not appropriate to directly use the charge density from this CV method (small perturbation) to represent the carrier density from DLCP measurement (large perturbation). In fact, the doping/charge densities are very likely to be skewed to larger values with the CV measurement if significant densities of deep trap states exist within the depletion region, as shown in previous studies[1]. Even in ref. 2, the experiment results from Heath et al. showed that the CV measurement actually overestimated the free carrier density in the $CuInSe_2$ solar cell compared to the DLCP results by several times.

Nevertheless, even under the framework of using carrier density and profile distance directly from CV measurement to represent those from DLCP measurement, the following analysis shows the geometry and diffusion capacitance do not impact the measuring of charge *trap* densities by DLCP method. Equation (1) predicts that carrier density is independent of applied ac frequency ($\omega$), however, the experimental results in ref. 3 clearly show a frequency dependent carrier distribution. As a matter of fact, the frequency dependent capacitance is a critically important characteristics of charge traps, which allows the determination of their density using the DLCP method. Without a frequency dependent carrier density, the traps would not show a trap energy resolution as shown in Fig. 3 in ref. 3. It should be noted that the trap density distributions shown in ref. 3 were derived from the difference of the carrier densities at different ac frequencies under the same dc biases. The calculated carrier or trap density would be zero by the DLCP method if there is no frequency dependent capacitance. In another word, the charge density derived from equation (1) does not show up as a background carrier density in the trap profiling derived from DLCP method, and thus does not dismiss the application of DLCP in characterizing the trap densities.

Equation (1) does not represent the capacitance of a semiconductor device, because it has an unreasonably constant junction capacitance of a p-n junction and also neglects the frequency dependence of the diffusion capacitance, though we understand the trapping/detrapping of charges induced capacitance was omitted from this equation on purpose. The dependence of the diffusion capacitance ($C_d$) on the dc bias ($V$) and ac frequency ($\omega$) of a typical $n^+$-$p$ junction is reported in textbook (ref. 4). As shown in the Supplementary materials section A1, this makes equation (1) applies only at low frequencies. At high frequencies, the contribution of diffusion capacitance to total capacitance reduces with a relationship of $C_d \propto \omega^{-0.5}$, which indicates the diffusion capacitance is less important at higher frequencies. Therefore, the contribution of diffusion capacitance to the measured doping (not trap) density is very small at high frequency. To verify whether the diffusion capacitance dominates the total capacitance at large forward biases in perovskite solar cells, we measured the *C-V* curves of a silicon diode (p-n junction) and perovskite single crystal and thin film solar cells at the ac frequency ($\omega$) of 62 kHz, and the result is shown in Fig. 1A. The *C-V* of the silicon diode clearly follows the relationship between $C_d$ and *V* for $V \geq$ 0.4V (detailed in the Supplementary materials section A1). While for the perovskite solar cell, the *C-V* behaviors at large forward biases do not follow equation (1). Since $C_d$ is related to both the forward dc bias *V* and $\omega$ (Supplementary materials section A1), we further checked whether *C* is dominated by $C_d$ at large forward biases over a wide range of $\omega$. The measured *C-$\omega$* curves and calculated $C_d$-$\omega$ plots are shown in Figs. 1B-D. The parameters used for the calculation of the $C_d$-$\omega$ are summarized in Table S1. Again, the calculated $C_d$-$\omega$ curves fit well for the Si diode in the low ac frequency region ($\omega\tau \ll 1$) where $C_d$ is independent of $\omega$ (Supplementary materials section A1), and $C_d$ is proportional to $\omega^{-1/2}$ at high ac frequencies ($\omega\tau \gg 1$) (Fig. 1B). In contrast, the calculated $C_d$ are 5 and 11 orders of magnitude smaller than the measured capacitance at large *V* over the wide range of $\omega$ for the thick single crystal and thin film perovskite solar cells, respectively (detailed in the Supplementary materials section A1). This is mainly attributed to the low minority carrier density in perovskites due to the self-doping of perovskite materials. Therefore, the $C_d$ should be negligible in perovskite solar cells within the dc bias range of 0-0.9 V adopted for the carrier density profiling in experiment. In this case, although the calculated carrier densities by CV method have a similar U-shape with that measured by DLCP, it requires a significantly larger forward dc bias of 1.0-1.6 V to complete the U-shape curve (Supplementary materials Fig. S1), which does not agree with the experimental data (0.2-0.9 V) in Ref. 3. The measured *C* at different dc biases and frequencies in Fig. 1C, D do not follow the predicted behavior of $C_d$ with $\omega$ by equation S1-3, either. The measured *C* keeps increasing with the decrease of $\omega$ in the low ac frequency region, instead of being a constant. We attribute this to the contribution of deep trap states from the junction capacitance. Moreover, even at the plateau region of *C-$\omega$* curves, C does not increase exponentially with *V*. These results further prove that the measured *C* for perovskite solar cells at forward dc biases is still dominated by junction capacitance and charge trapping/detrapping capacitance, while any further analysis based on equation (1) would be inaccurate without taking the junction capacitance into consideration especially when the doping/trap distributions are not uniform in the perovskite.

Finally, since we may also use DLCP to evaluate doping density distribution in semiconductor devices by conducting the measurement at high frequency. An absence of subtraction process impose a question whether the measured carrier density is right. We recently show that the carrier density measured by DLCP in the perovskite films with thickness of 1 μm changed after aging, and had a lower carrier density of ~ $3\times10^{14}$ cm$^{-3}$ than what calculated in the commentary [5], which indicates the diffusion capacitance is still not dominating either. Here we estimated the possible contribution of the diffusion capacitance caused "interfacial carrier density" to the DLCP measured carrier densities in perovskite solar cells by treating the diffusion capacitance in the same way as we did in DLCP measurement (detailed in Supplementary materials section A2 and Figs. S2, S3). The contribution of diffusion capacitance is clearly negligible due to the high frequency of measurement and small minority carrier density, indicating that the measured interfacial charge densities in perovskite solar cells are not caused by the charge injections. In addition, the influence of the $C_g$ on the carrier density measurement is prominent only when the applied bias is more than enough to fully deplete the perovskite layers, which was actually excluded in the carrier density profiling by DLCP measurement (Supplementary materials section A3 and Figs. S4, S5).

**Supplementary Materials**
Table S1
Fig. S1-S5
Section A1-A3

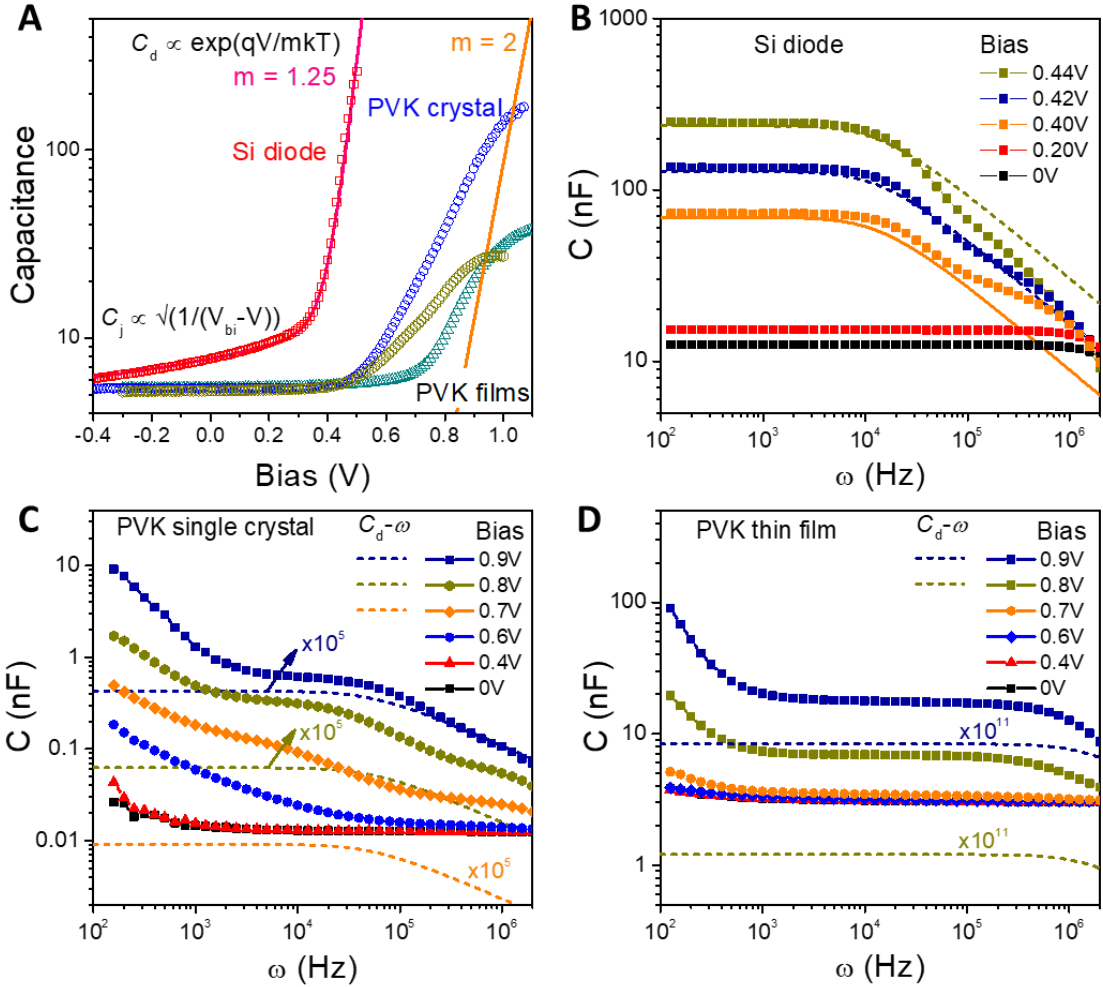

**Fig. 1. Contribution of the diffusion capacitance to the total capacitances.** (**A**) Normalized *C-V* curves of a silicon (Si) diode (red), perovskite thin single crystal (PVK crystal) (blue) and polycrystalline thin film (PVK films) (green and khaki) solar cells measured at ac frequency (*ω*) of 62 kHz. The solid red and orange lines plot the dependence of $C_d$ on exp(qV/mkT) with m = 1.25 and 2, respectively. *C-ω* curves of the (**B**) Si diode, (**C**) perovskite thin single crystal and (**D**) perovskite thin film solar cells measured at difference dc biases. The dashed lines show the calculated $C_d$-$\omega$ curves with equation $C_d = \frac{Aq^2 n_{p0} L_n}{kT\tau_n} \exp\left(\frac{qV}{mkT}\right) (((1+\omega^2\tau_n^2)^{\frac{1}{2}} - 1)/2\omega^2)^{1/2}$ at the corresponding dc biases.

Supplementary Materials for
**Response to comment on "Resolving spatial and energetic distributions of trap states in metal halide perovskite solar cells"**
Zhenyi Ni, Shuang Xu and Jinsong Huang*

Department of Applied Physical Sciences, University of North Carolina, Chapel Hill, NC 27599, USA.

*Corresponding author. Email: jhuang@unc.edu

**Table S1** Parameters used for the calculation of the diffusion capacitance.

| Materials | $N_C, N_V$ | $E_g$ | $n_i$ | $n_{p0}$ | $L_n$ | $\tau_n$ | $m$ |
|---|---|---|---|---|---|---|---|
| Silicon | | 1.12 eV | 1E10 cm$^{-3}$ | 2E5 cm$^{-3}$ | 450 μm | 120 μs | 1.25 |
| Perovskite crystal | 2E18 cm$^{-3}$ | 1.55 eV | 1.8E5 cm$^{-3}$ | 3 cm$^{-3}$ | 10 μm | 30 μs | 2 |
| Perovskite thin film | 2E18 cm$^{-3}$ | 1.55 eV | 1.8E5 cm$^{-3}$ | 3E-5 cm$^{-3}$ | 1 μm | 1 μs | 2 |

**Discussion of the fitting parameters:**

**Effective density of states of conduction and valence bands ($N_C$, $N_V$):** the effective density of states for the conduction and valence band of perovskite were chosen from the same reference used in the commentary[1].

**Intrinsic carrier density ($n_i$):** the $n_i$ of silicon was chosen from Ref. 2. The $n_i$ of perovskite is calculated by using $n_i = (N_C N_V)^{1/2} \exp(-E_g/2kT)$ at $T$ = 300K. Though divergence may exist between the calculated and measured $n_i$, the difference of the values is basically within one order of magnitude for most semiconductors[2].

**Minority carrier density ($n_{p0}$):** for the silicon diode, the minority of the weakly doped side of the junction was calculated by $n_{p0} = n_i^2/p_{p0}$, where the majority carrier density $p_{p0}$ was estimated from the Mott-Schottky plot of the silicon diode. For perovskite devices, the doping (majority carrier) density measured in MAPbI$_3$ single crystals by Hall Effect measurement was reported to be around $10^{10}$ cm$^{-3}$,[3] and that measured in polycrystalline thin films by Hall Effect measurement was in the order of $10^{15}$ cm$^{-3}$.[4,5] The $n_{p0}$ in perovskite single crystal and thin films are calculated to be 3 cm$^{-3}$ and 3×10$^{-5}$ cm$^{-3}$, respectively.

**Diffusion length of the minority ($L_n$):** the $L_n$ of silicon was read from Ref. 6. The $L_n$ of thin perovskite single crystals was chosen from Ref. 7, which was around 10 μm. Most reported $L_n$ of perovskite thin films were less than one micrometer. A representative value of 1 μm was chosen for the calculation of diffusion capacitances.

**Minority carrier lifetime ($\tau_n$):** the minority carrier lifetime does not affect the calculation of the carrier density at low frequencies ($\omega \tau_n \ll 1$), but determines the demarcation frequency when the diffusion capacitance starts to drop with the further increase of $\omega$. For silicon, the $\tau_n$ was chosen to be 120 μs which generates the best fitting results and also similar to the value listed in Ref. 2. For perovskite thin single crystal and thin films, $\tau_n$ were chosen to be close to the reported values in Refs. 7-8.

**Diode ideality factor ($m$)**: the ideality factor was chosen in the range from 1 to 2. For the silicon diode, $m = 1.25$ was chosen to give the best fitting of the measured $C$-$V$ plot. For perovskite devices, $m = 2$ was chosen to calculate the diffusion capacitances.

### A1. Dependence of diffusion capacitance on dc bias and ac frequency

The dependence of the diffusion capacitance ($C_d$) on the dc bias ($V$) and ac frequency ($\omega$) of a typical $n^+$-$p$ junction is derived from Ref. 2 by extending the dependence of the small-signal current density on voltage, for which the imaginary part gives the $C_d$:

$$C_d = \frac{Aq^2 n_{p0} L_n}{kT\tau_n} \exp\left(\frac{qV}{mkT}\right) \sqrt{\frac{\sqrt{1+\omega^2\tau_n^2}-1}{2\omega^2}}, \tag{S1}$$

where $A$ is junction area, $q$ is the elementary charge, $n_{p0}$ is the minority (electron) carrier density of the $p$-type layer, $L_n$ is the diffusion length of the minority in the $p$-type layer, $k$ is the Boltzmann's constant, $T$ is the temperature, $\tau_n$ is the minority carrier lifetime in the $p$-type layer, $m$ is the ideality factor which ranges from 1 to 2 in all optimized perovskite solar cells.

At low frequencies ($\omega\tau_n \ll 1$), $\sqrt{1+\omega^2\tau_n^2} \approx 1 + \frac{\omega^2\tau_n^2}{2}$, $C_d$ is approximated to

$$C_d = \frac{Aq^2 n_{p0} L_n}{2kT} \exp\left(\frac{qV}{mkT}\right), \tag{S2}$$

which is independent of frequency $\omega$. At high frequencies ($\omega\tau_n \gg 1$), $C_d$ is given by

$$C_d = \frac{Aq^2 n_{p0} L_n}{kT} \exp\left(\frac{qV}{mkT}\right) (2\omega\tau_n)^{-1/2}. \tag{S3}$$

As shown in Fig 1A, for the silicon diode, two regions with different bias ranges where the junction capacitance ($C_j$) and $C_d$ dominant respectively can be clearly distinguished. The $C_d$-$V$ at the large bias region is well fitted by $C_d \propto \exp(qV/mkT)$ with $m = 1.25$. For perovskite solar cells, the $C$-$V$ behaviors at large forward biases do not follow the characteristic of $C_d$ on $V$ with $m$ up to 2, indicating $C_d$ is not dominating the total capacitance even at high bias up to $V_{OC}$.

The $C_d$-$\omega$ curves of the silicon diode and perovskite solar cells are calculated by using equation (S1) with the parameters listed in Table S1. In the low frequency region, $C_d$ is basically independent of $\omega$, which is given by equation S2. In the high frequency region, $C_d$ monotonously decreases with the increase of $\omega$, following the relationship given by equation S3. For perovskite solar cells, the calculated $C_d$ for perovskite single crystal and thin film solar cells at the low ac frequencies ($\omega\tau_n \ll 1$) and $V = 0.9$ V with the parameters listed in Table S1 are about $4\times10^{-6}$ nF and $8\times10^{-11}$ nF, respectively, which are at 5 and 11 orders of magnitude smaller than the measured $C$ under the same conditions. Therefore, the influence of $C_d$ on the total capacitance measured at large forward dc biases is negligible for perovskite solar cells.

### A2. Influence of the diffusion capacitance on the DLCP carrier doping density

Here we treated diffusion capacitance in the same way as what we did in DLCP measurement to find out how the diffusion capacitance impact the carrier doping concentration in DCLP measurement. In DLCP measurement, a series of changing ac biases ($\delta V$) are applied to the junction. The carrier density including doping and trap density is derived from the coefficients $C_0$ and $C_1$ from the dependence of the total $C$ on $\delta V$:

$$C = C_0 + C_1 \delta V + C_2 \delta V^2 \ldots \tag{S4}$$

$$N_{DLCP} = -\frac{C_0^3}{2A^2 q \varepsilon_r \varepsilon_0 C_1} \tag{S5}$$

Now we discuss how $C_d$ would affect $C_0$ and $C_1$ at larger forward dc biases during the DLCP measurement. In DLCP measurement, $\delta V$ refers to the peak-to-peak value of the ac biases, and for each $\delta V$ applied a compensatory dc bias of $-1/2\delta V$ is added to the total bias to make the maximum applied forward voltage constant and fix the measurement region (Ref. 9). If the $C_d$ is not dependent on ac biases, as given by equation S1, the dependence of $C_d$ on $\delta V$ at a given forward dc bias of $V_0$ is

$$C_d = C' \exp\left(\frac{q\left(V_0 - \frac{1}{2}\delta V\right)}{mkT}\right) \tag{S6}$$

$$= C' \exp\left(\frac{qV_0}{mkT}\right)\left[1 - \frac{q}{2mkT}\delta V + \frac{q^2}{8m^2k^2T^2}\delta V^2 \ldots\right] \tag{S7}$$

where $C' = \frac{Aq^2 n_{p0} L_n}{2kT}$ or $\frac{Aq^2 n_{p0} L_n}{2kT}\left(\frac{2}{\omega \tau_n}\right)^{1/2}$ when $\omega\tau_n \ll 1$ or $\omega\tau_n \gg 1$, respectively.

Writing the $C_d$ in the form of $C_d = C_{d0} + C_{d1}\delta V + C_{d2}\delta V^2 \ldots$, we can obtain:

$$C_{d0} = C' \exp\left(\frac{qV_0}{mkT}\right), \tag{S8}$$

$$C_{d1} = -\frac{q}{2mkT} C' \exp\left(\frac{qV_0}{mkT}\right). \tag{S9}$$

Comparing equations S8 and S9, we can obtain $\frac{C_{d1}}{C_{d0}} = -\frac{q}{2mkT} \approx -\frac{19.4}{m}$ (V$^{-1}$) at $T = 300$K.

For the Si diode measured at the dc bias of 0.4 V and ac frequency of 62 kHz, the numerical values of $C_0$ and $C_1$ derived from the polynomial fitting of $C$ on $\delta V$ are 9.02e-8 and -1.34e-6, respectively (Fig. S2A), which agrees with the relationship between $C_{d0}$ and $C_{d1}$ with m = 1.29, demonstrating a typical characteristic of the diffusion capacitance at this dc bias. The derived $m$ value is consistent with that obtained from the $C$-$V$ and $C$-$\omega$ plots of the Si diode at low ac frequencies. At a high ac frequency of 1.8 MHz, the ratio of the numerical values of $C_1/C_0$ was around -2.2 at the dc bias of 0.4 V (Fig. S3A), which does not agree with the ratio of $C_{d1}/C_{d0}$ with $m$ up to 2. This indicates that $C_d$ should no longer dominate the measured $C$ at the high ac frequency regime for the Si diode, which is again reasonable. This analysis shows that the frequency dependent capacitance should not be ignored in calculating the total capacitance.

For perovskite single crystal and polycrystalline thin film solar cells, the derived $C_1/C_0$ ratios at the dc bias of 0.9 V and the ac frequency of 62 kHz are about -4.9 and -3.9, respectively (Figs. S2B, C), which significantly differ from the ratio of $C_{d1}/C_{d0}$ with $m$ up to 2 which is about -10. Here we have to consider the contribution of junction capacitance ($C_j$) to explain the significant different $C_1/C_0$ ratios at a large dc bias of 0.9 V:

$$C = C_j + C_d \tag{S10}$$

Then we have $C_0 = C_{j0} + C_{d0}$, $C_1 = C_{j1} + C_{d1}$. For $m = 2$, $C_{d1} = -10 C_{d0}$. Since the measured $C_1/C_0 > -5$, substituting $C_{d1}$ with $-10 C_{d0}$ and combining $C_{j1} < 0$, we can have $C_0 > 2 C_{d0}$. Writing $C_0 = 2C_{d0} + \Delta C$ where $\Delta C > 0$, the total ($N_{tot}$) and diffusion-induced ($N_d$) carrier densities can be estimated by:

$$N_{tot} = -\frac{C_0^3}{2A^2 q \varepsilon_r \varepsilon_0 C_1} > \frac{C_0^2}{10 A^2 q \varepsilon_r \varepsilon_0} = \frac{(2C_{d0} + \Delta C)^2}{10 A^2 q \varepsilon_r \varepsilon_0}, \tag{S11}$$

$$N_d = -\frac{C_{d0}^3}{2A^2 q\varepsilon_r\varepsilon_0 C_{d1}} = \frac{C_{d0}^2}{20A^2 q\varepsilon_r\varepsilon_0}. \tag{S12}$$

As a result, $N_{tot}$ is at least 8 times larger than $N_d$ for the perovskite solar cells at the dc bias of 0.9 V and at frequency of 62 kHz (corresponding the interface), *even without considering the theoretical limitation of $C'$ by using the parameters listed in Table S1*. A quick way to check the possible maximum contribution of $N_d$ to $N_{tot}$ is by calculating $(-C_1/10C_0)^3$ at different forward dc biases. At the high ac frequency of 1.8 MHz, the derived $C_1/C_0$ ratios for the perovskite thin single crystal and polycrystalline thin film solar cells at the dc bias of 0.9 V were -3.8 and -2.7, respectively (Figs. S3B, C). As a result, the interfacial $N_{tot}$ is at least 17 and 50 times larger than $N_d$ for the perovskite single crystal and thin film solar cells respectively at the high ac frequency of 1.8 MHz (Figs. S4C, S5C). Once the actual $C'$ in the real perovskite devices is adopted (section A1), $C_{d0}$ would be orders of magnitude smaller than $C_{j0}$, making the influence of the charge injections to the interfacial carrier/doping densities really negligible.

### A3. Influence of the geometrical capacitance on the DLCP carrier doping density

In DLCP measurement, the applied dc bias sweep the depletion region from one interface toward the other to scan the charge and trap distribution. There is a critical dc bias to fully deplete the perovskite layer. When the applied bias is larger than this critical value, $N(w) = -2(dC^{-2}/dV)^{-1}/q\varepsilon_r\varepsilon_0$ tells the actual doping concentration when there is no trapping based on the Schottky-Mott analysis. When the applied bias is smaller than this critical value, the capacitance does not change anymore, which equals to the geometrical capacitance $C_g$ and thus $dC^{-2}/dV$ is almost zero, which gives the huge carrier density in the commentary. We excluded the data for the biases less than the critical value, because when the depletion region reaches the metal/perovskite interface, there is no point to increase the reverse bias to further push the depletion region edge.

Since $C_g$ is independent of dc biases, the coefficients of the high order terms of $\delta V$ ($C_1$, $C_2$…) in equation S4 should be infinitely small or zero when the $C_g$ dominates the total $C$. This would causes significant variation of $C_0$, and $C_1$ from the fitting of $C$ on $\delta V$, thus leading to invalid carrier densities calculated (Figs. S4, S5). In our measurement, we carefully excluded this situation to occur, ruling out the influence of the $C_g$ on the measured carrier density distributions in perovskite solar cells.

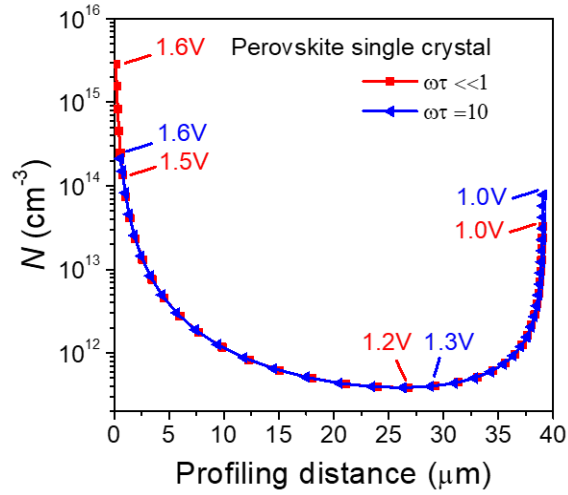

**Fig. S1.** CV method calculated carrier density distributions in a perovskite single crystal solar cell based on the assumption $C = C_g + C_0 \exp(qV/mkT)$ at low ac frequency ($\omega\tau \ll 1$) and high ac frequency ($\omega\tau = 10$). The dc biases needed to achieve the corresponding carrier densities are denoted close to the curves. In experiment of Ref. 10, the adopted bias range for the DLCP carrier density profiling was 0.2 - 0.9 V, which is very different from the bias here.

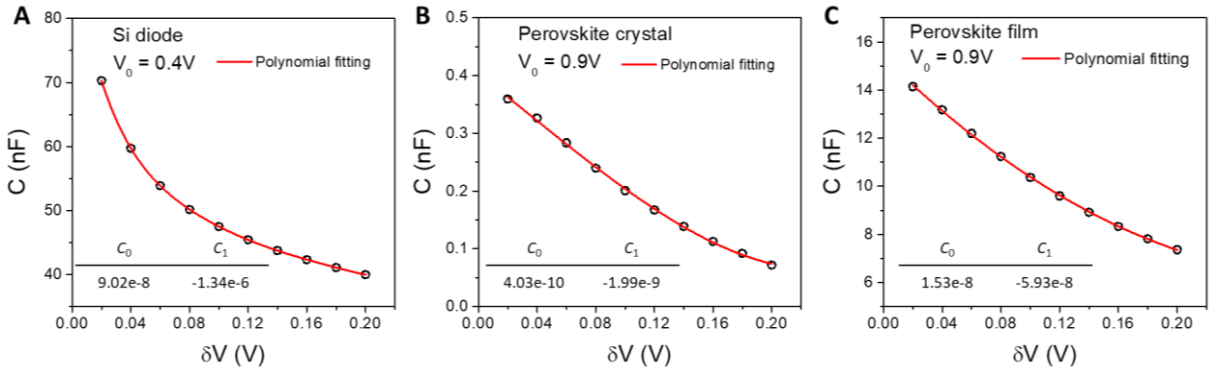

**Fig. S2. Influence of the diffusion capacitance at large forward bias and low frequency.** Dependence of the capacitance on the ac bias $\delta V$ at a large forward dc bias ($V_0$) for the (A) silicon diode, (B) perovskite thin single crystal solar cell and (C) perovskite thin film solar cell measured at the ac frequency $\omega$ of 62 kHz. The numerical values of $C_0$ and $C_1$ derived from the polynomial fitting of $C$-$\delta V$ with $C = C_0 + C_1\delta V + C_2\delta V^2$…are listed inside the figure.

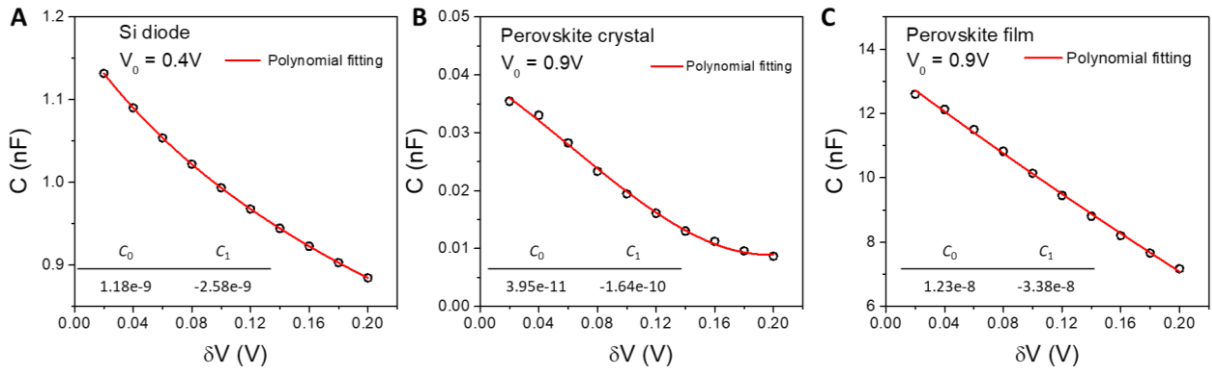

**Fig. S3. Influence of the diffusion capacitance at large forward bias and high frequency.** Dependence of the capacitance on the ac bias $\delta V$ at a large forward dc bias ($V_0$) for the (A) silicon diode, (B) perovskite thin single crystal solar cell and (C) perovskite thin film solar cell measured at the ac frequency $\omega$ of 1.8 MHz. The numerical values of $C_0$ and $C_1$ derived from the polynomial fitting of $C$-$\delta V$ with $C = C_0 + C_1\delta V + C_2\delta V^2$…are listed inside the figure.

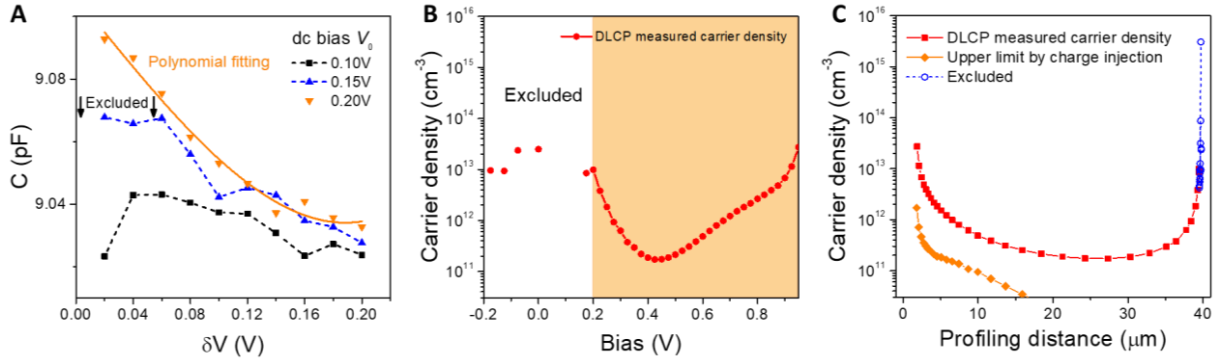

**Fig. S4. DLCP measurement of a perovskite single crystal solar cell at high ac frequency.** (A) Dependence of the capacitance on the ac bias $\delta V$ at small forward dc biases ($V_0$) for a perovskite thin single crystal solar cell measured at the ac frequency $\omega$ of 1.8 MHz. When the bias is less than 0.2 V, the $C$ - $\delta V$ showed an irregular shape, resulting in invalid fitting for $C_0$ and $C_1$. (B) Dependence of the carrier densities on the dc bias measured by DLCP at $\omega$ of 1.8 MHz. (C) DLCP measured carrier density distribution in the perovskite single crystal solar cell at $\omega$ of 1.8 MHz. Those invalid data point measured at the dc biases smaller than 0.2 V are marked by blue circles. The upper limit carrier density by charge injections is derived from the analysis detailed in section A2.

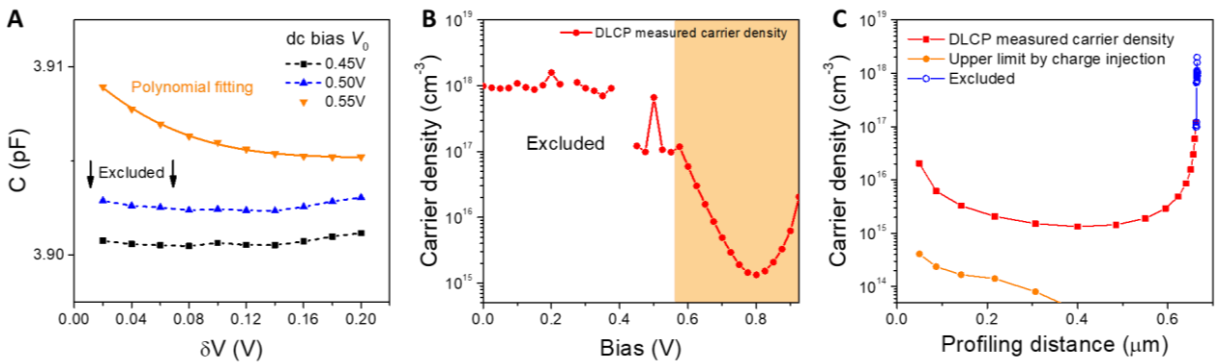

**Fig. S5. DLCP measurement of a perovskite thin film solar cell at high ac frequency.** (A) Dependence of the capacitance on the ac bias $\delta V$ at small forward dc biases ($V_0$) for a perovskite thin film solar cell measured at the ac frequency $\omega$ of 1.8 MHz. When the bias is less than 0.55 V, the $C$ - $\delta V$ showed an irregular shape, resulting in invalid fitting for $C_0$ and $C_1$. (B) Dependence of the carrier densities on the dc bias measured by DLCP at $\omega$ of 1.8 MHz. (C) DLCP measured carrier density distribution in the perovskite thin film solar cell at $\omega$ of 1.8 MHz. Those invalid data point measured at the dc biases smaller than 0.5V are marked by blue circles. The upper limit carrier density by charge injections is derived from the analysis detailed in section A2.